\documentclass[aps,twocolumn,a4paper]{revtex4}
\usepackage{amsmath,graphicx}

\begin{document}

\title{An integrated atom-photon junction}

\author{M. Kohnen}\author{M. Succo}\author{P. G. Petrov}\author{R. A. Nyman}\author{M. Trupke}\author{E. A. Hinds}

\affiliation{Centre for Cold Matter, Blackett Laboratory, Imperial College London, Prince Consort Road, SW7 2BW, United Kingdom}

\date{\today}

\begin{abstract}
Photonic chips that integrate guides, switches, gratings and other components, process vast amounts of information rapidly on a single device. A new branch of this technology becomes possible if the light is coupled to cold atoms in a junction of small enough cross section, so that small numbers of photons interact appreciably with the atoms. Cold atoms are among the most sensitive of metrological tools and their quantum nature also provides a basis for new information processing methods. Here we demonstrate a photonic chip which provides multiple microscopic junctions between atoms and photons. We use the absorption of light at a junction to reveal the presence of one atom on average. Conversely, we use the atoms to probe the intensity and polarisation of the light. Our device paves the way for a new type of chip with interconnected circuits of atoms and photons.
\end{abstract}

\maketitle

Micro-fabricated chips are widely used to control clouds of ultra-cold atoms and Bose-Einstein condensates \cite{Fortagh07,ReichelVuletic}. Recently, the idea has been extended to the control of ions \cite{Seidelin06} and similar possibilities exist for molecules \cite{Andre06}. This atom-chip technology provides a way to miniaturise existing atomic physics devices. In addition, it promises new devices that take advantage of the elementary quantum nature of atoms \cite{Jaksch99,Chen06, Gleyzes09}, ions \cite{CiracZoller95} and molecules \cite{DeMille02,Andre06}, together with photons \cite {Politi08}. Although laser light is an essential tool for preparing, controlling and interrogating the atomic systems on a chip, the integration of optics into atom chips is still in its infancy. Millimetre-sized vapour cells have enabled optical spectroscopy \cite{Yang07}, clocks \cite{Knappe04} and magnetometry \cite{Knappe04} on a chip, and etched mirrors on a silicon wafer have been used to integrate magneto-optical traps \cite{Pollock09}. On a scale ten times smaller, several groups have explored how optical fibres, typically $125\,\mu$m in diameter, may be glued \cite{Quinto-Su04, Eriksson05,Takamizawa06} or otherwise attached \cite{Wilzbach09} to a chip. A pair of these fibres looking into each other can be used to detect an atom cloud and can reach close to single atom sensitivity \cite{Eriksson05}.  When reflective coatings are added, the gap between two fibres \cite{Horak03,Colombe07} or between one fibre and a micro-fabricated mirror \cite{Trupke07} becomes a Fabry-Perot resonator. Similarly, a fibre can be coupled to a micro-disk resonator \cite{Dayan08,Aoki09}. These devices can achieve strong atom-photon coupling for applications in quantum information processing.

\begin{figure}[t]  
		\centering
        \includegraphics[width=6cm]{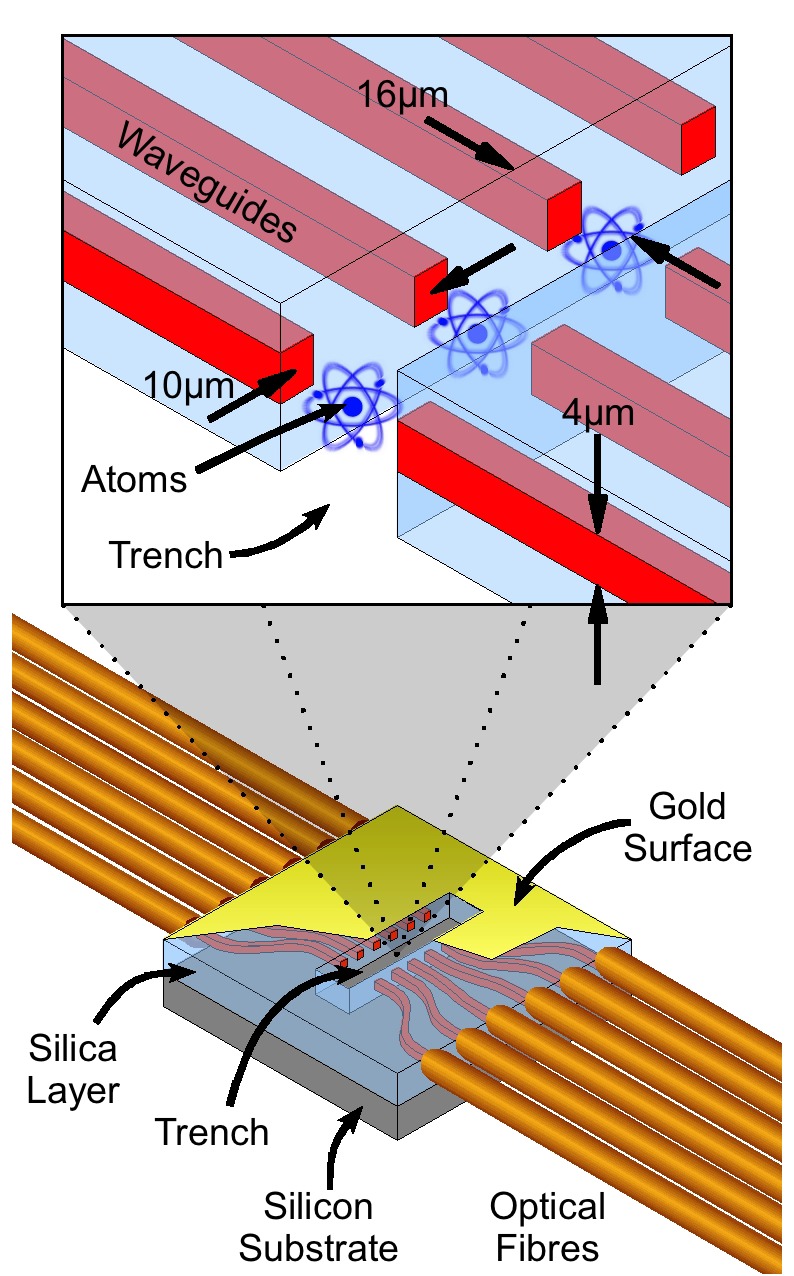}
        \caption{{\bf Schematic diagram of the integrated-waveguide atom chip.} A silicon substrate supports a layer of silica cladding, within which $4\mu$m-square doped silica waveguide cores are embedded. There are 12 parallel waveguides (for clarity, only six are shown) spaced at the centre of the chip by $10\mu$m. These flare out at the edges of the chip so that optical fibres can be connected. A $22\mu$m-deep trench cutting across the waveguides is narrow enough ($16\mu$m) that 65\% of the light entering the trench is collected by the waveguides on the far side in the absence of atoms. An atom in the trench affects the phase and the intensity of the transmitted light. Conversely, the light affects the state of the atom. Thus each waveguide provides a microscopic atom-photon junction. The top layer of the chip is coated with gold to reflect laser light used for cooling the atoms. Current-carrying wires below the chip provide magnetic fields to trap and move the atoms.}
        \label{Fig: chip diagram}
\end{figure}

This manuscript reports a further order of magnitude scale reduction, in which the bulky optical fibre is replaced by an array of twelve integrated optical waveguides, spaced on a pitch of only $10\,\mu$m and having a $4\,\mu$m square core, as illustrated in Figure~\ref{Fig: chip diagram}. The array is cut by a trench, where trapped atomic particles can interact with the light. The optical interaction regions are so small that a single atom obscures a significant fraction of the light, and there can be many such regions closely spaced. By demonstrating that these micro-fabricated atom-photon junctions are practical, we pave the way for integrated devices that exploit the quantum properties of both atoms and photons.
\clearpage

For this first demonstration, we have released cold atoms into a junction to measure its sensitivity and to demonstrate the basics of its operation. Every few seconds, $^{87}$Rb atoms are cooled and collected from a room-temperature vapour by a Low-Velocity Intense Source (LVIS), then transferred to a magneto-optical trap (MOT) 3\,mm from the chip surface, where the atom density is up to $10^{-2}$ atoms/$\mu$m$^3$ and the temperature is $\sim100 \mu$K. We push this cloud towards the chip just before switching off the MOT light and magnetic field, thereby launching the atoms at $40\,$cm/s into the trench. The light beams from the waveguides diverge only slightly as they cross the trench, with $w$ - the 1/e radius of the field - growing from $w_0 = 2.2\,\mu$m to $w = 2.8\,\mu$m. Since the width of the trench is $L=16\,\mu$m, any given beam interacts with roughly one atom of the cloud as it passes through. Each atom crosses the light in $\sim7\,\mu$s, scattering up to $130$ photons (the fully saturated rate is $\Gamma/2=1.9\times 10^7\,$s$^{-1}$).

\begin{figure}
        \centering
        \includegraphics[width=5cm]{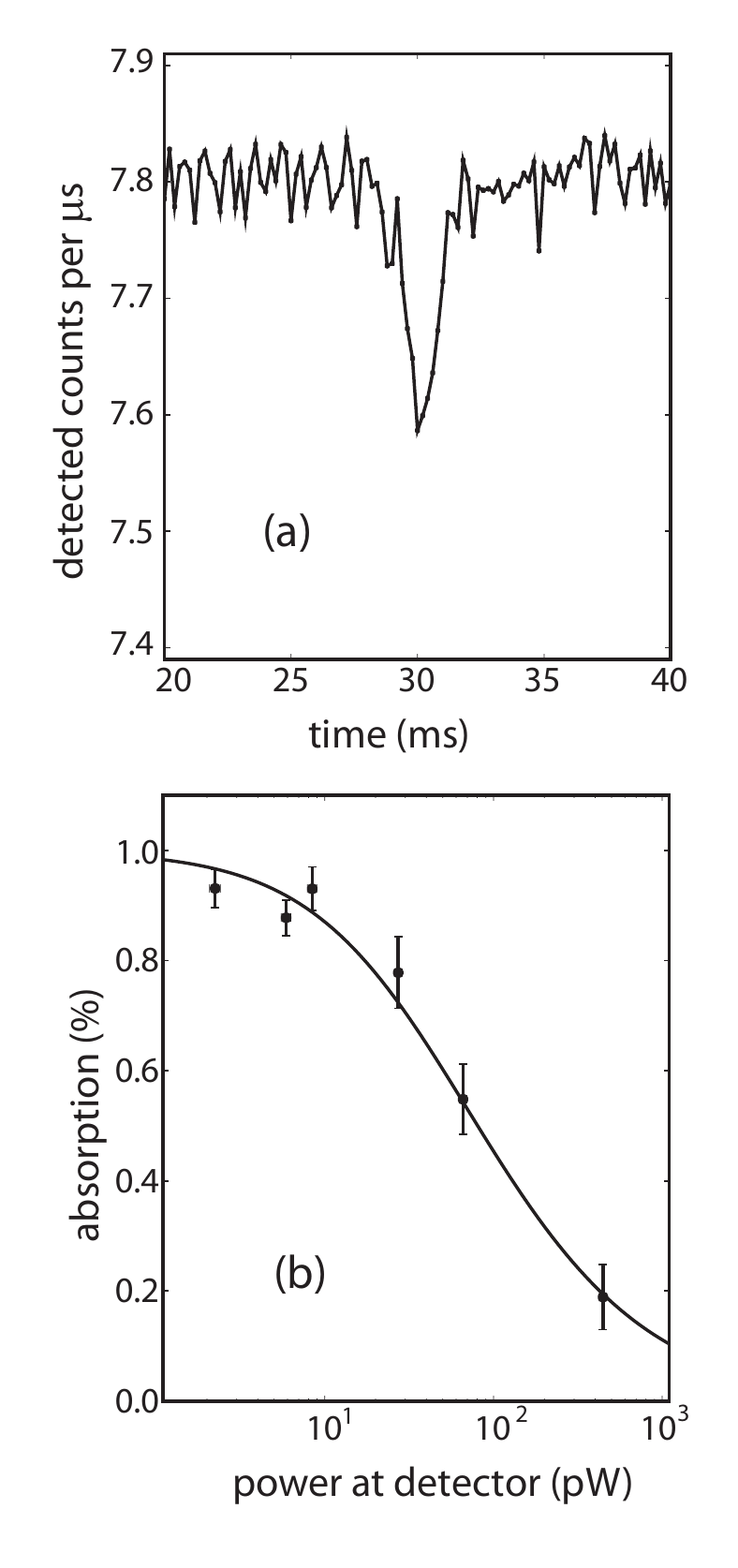}
        \caption{{\bf Absorption of light by atoms in the trench.} (a): Laser power transmitted through a waveguide pair (Fig.~1) to a photon-counting avalanche photodiode (APD) detector. There are $7.8\times10^6$ counts per second, corresponding to $5$~pW of light power (quantum efficiency: 50\%, dead time: 32~ns). The 2.6\,\% dip indicates $8\times 10^{-3}$ atoms/$\mu$m$^{3}$ in the trench, corresponding to just one atom on average in the illuminated volume of $\sim 100\,\mu$m$^3$. The width of the dip corresponds to our cloud size of slightly less than $1~$mm. (b): Measured fraction of light absorbed by the atoms (with 2~ms integration) versus detected power. The solid line is a theoretical curve taking into account the variation of light intensity at different positions in the trench and the optical pumping of the atoms, which affects $I_{sat}$ by polarising them (see Methods section \ref{sec: Isat}). There are two fitting parameters: the atom number density, which sets the scale of the absorption, and the ratio of light power in the trench to that at the APD, which re-scales the theory along the power axis.
        We see good agreement between experiment and theory, which indicates that we understand the interaction between the atoms and the light in the trench. The fit tells us that the power in the trench is six times that at the detector, which is consistent with the known transmission of 65\% across the trench and 25\% through the fibre couplers on the way to the detector. Error bars give 1$\sigma$ statistical uncertainties for Poissonian light with detector dead time. }
        \label{Fig: absorption}
\end{figure}

With low-intensity resonant light in the trench at frequency $\omega = 2\pi\times 385$~THz, and rubidium atoms of number density $\rho$, we expect the absorbed fraction of the light to be given by
\begin{equation}
f=\frac{\hbar \omega \Gamma L}{2 I_{sat}} \rho=3.2\,\mu\rm{m}^{3}\times\rho.
\end{equation}
The saturation intensity $I_{sat}$ depends on the polarisation of the atoms and the light, and for our first experiments it is well approximated by $I_{sat}=22\,$pW/$\mu$m$^2$ (see Methods section \ref{sec: Isat}). Thus, the atoms in our experiment should absorb up to 3\% of the light. The absorption that we observe, as shown in Fig.~\ref{Fig: absorption}(a), is entirely consistent with this prediction. This dip is the first signal from a fully integrated, microscopic atom-photon junction.

If the intensity in the trench $I$ approaches $I_{sat}$, the scattering rate begins to saturate and the absorbed fraction decreases according to the relation
\begin{equation}
f(I)=\frac{f(I\rightarrow 0)}{1+I/I_{sat}}
\end{equation}
The bleaching of the atoms in this way lets us measure the absolute intensity of the light in the trench. Figure~\ref{Fig: absorption}(b) shows the measured decrease of absorption as the detected power is increased, together with a fitted theory curve. The fit relates intensity in the trench to power at the detector and demonstrates that the latter is six times smaller than the power in the trench, as expected (see Methods A).  These experiments show that the junction works: the light detects the atoms, the atoms sense the light, and both perform in agreement with theory.

For some applications it will be important to control the polarisation of the light in the trench as well as its intensity. We find that light prepared with positive helicity, $\xi^+$, and delivered to the chip through one of the optical fibres, still has typically 87\% of the power in the $\xi^+$ mode when it emerges from the corresponding output fibre, provided the fibres are securely held so that they cannot move in the ambient currents of air \cite{footnote1}. This suggests that the light in the trench is polarised, but we were also able to check the polarisation directly, using the atoms themselves as the probe.

\begin{figure}[ht]
        \centering
       \includegraphics[width=8cm]{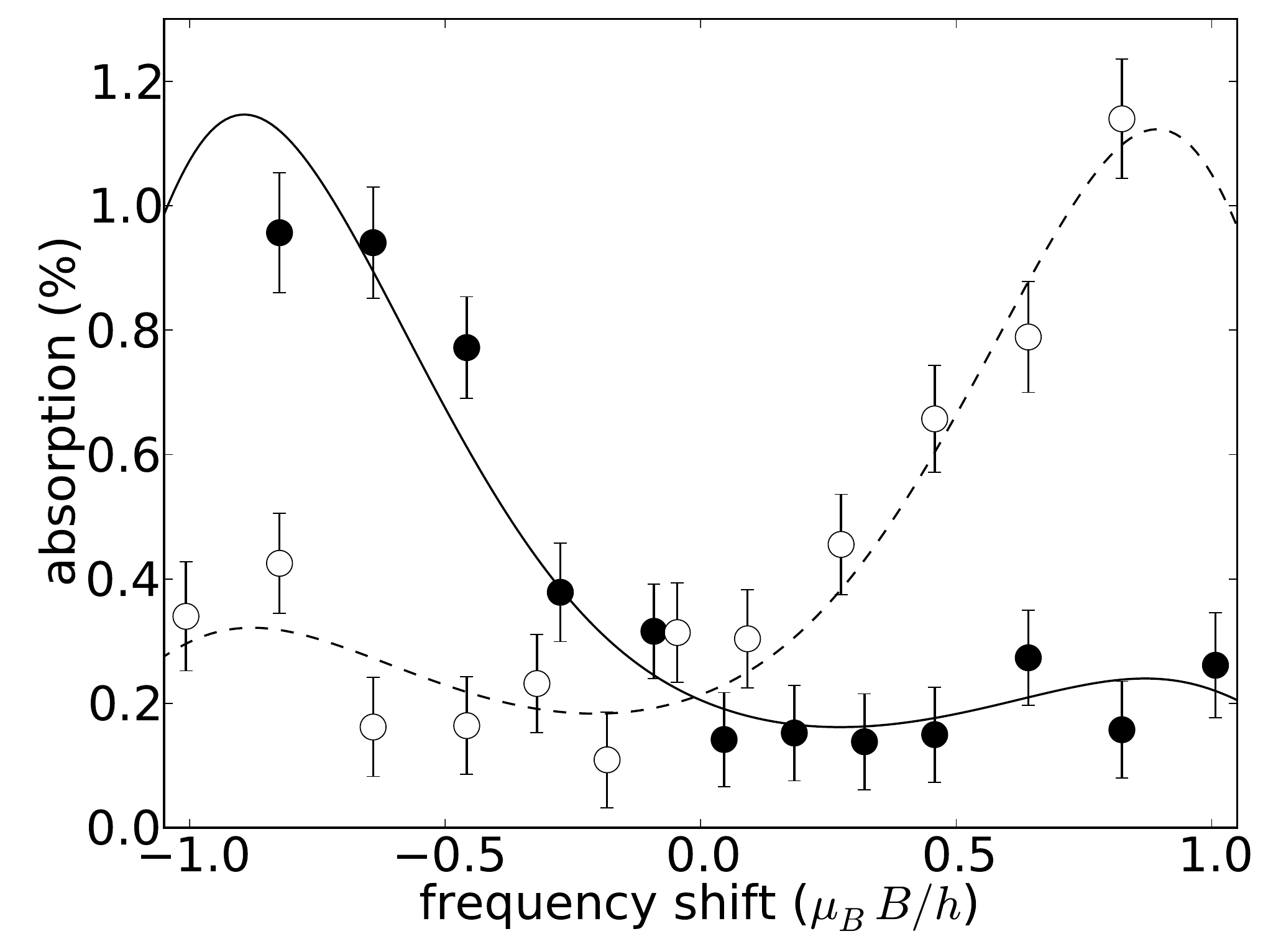}
		\caption{{\bf Zeeman-split absorption spectrum measures the light polarisation.} $D_2$ absorption spectrum of $^{87}$Rb, measured on the hyperfine component $\left|F=2\right>\rightarrow\left|F'=3\right>$ with a magnetic field of 0.78\,mT parallel to the waveguides. When pure $\xi^-$ light (filled circles) is coupled into the input fibre the Zeeman shift is almost $-\mu_B B/h$. The line is a theory obtained by summing over all the Zeeman components of the spectrum, fitted to the data with two parameters: the over-all normalisation and the ratio of $\sigma^+$ to $\sigma^-$ intensity. This fit tells us that $(85\pm3)\%$ of the power in the trench is still in the $\xi^-$ mode. For $\xi^+$ light (open circles) the shifts are opposite. The fit to these points gives $(81\pm3)\%$ of power in the $\xi^+$ mode. The experiment was done at sufficiently low light intensity that optical pumping was negligible, allowing the theory to take equal populations of the $m_F$ levels. Error bars give 1$\sigma$ statistical uncertainties for Poissonian light with detector dead time.}
        \label{Fig: zeeman}
\end{figure}

Figure~\ref{Fig: zeeman} shows the absorption of the light versus frequency in the presence of a 0.78\,mT magnetic field parallel to the waveguides. This spectrum indicates the ratio of $\sigma^+$ to $\sigma^-$ power in the trench, where $\sigma^+$($\sigma^-$) means the angular momentum of the light is parallel (antiparallel) to the magnetic field. The light excites the $^{87}$Rb $D_2$ transition $\left|F=2,m_F\right>\rightarrow\left|F'=3,m_F'\right>$, for which the Zeeman shift of $\sigma^\pm$ transitions is $\frac{1}{6}(m_F\pm4)\mu_B B/h$. The strongest excitation with $\sigma^+(\sigma^-)$ light comes from $m_F=+2(-2)$ and is shifted by $+(-)\mu_B B/h$. With $\xi^-$ input light (filled circles) the strong peak is close to $-\mu_B B / h$, indicating that the power in the trench is still mainly of $\xi^-$ polarisation. When we reverse the helicity (open circles) the shift reverses. We conclude that the polarisation in the trench can be controlled by adjusting the input light to have anything up to $90\%$ of the power in pure $\sigma^+$ or $\sigma^-$ polarisation.

\begin{figure}
        \centering
      \includegraphics[width=8cm]{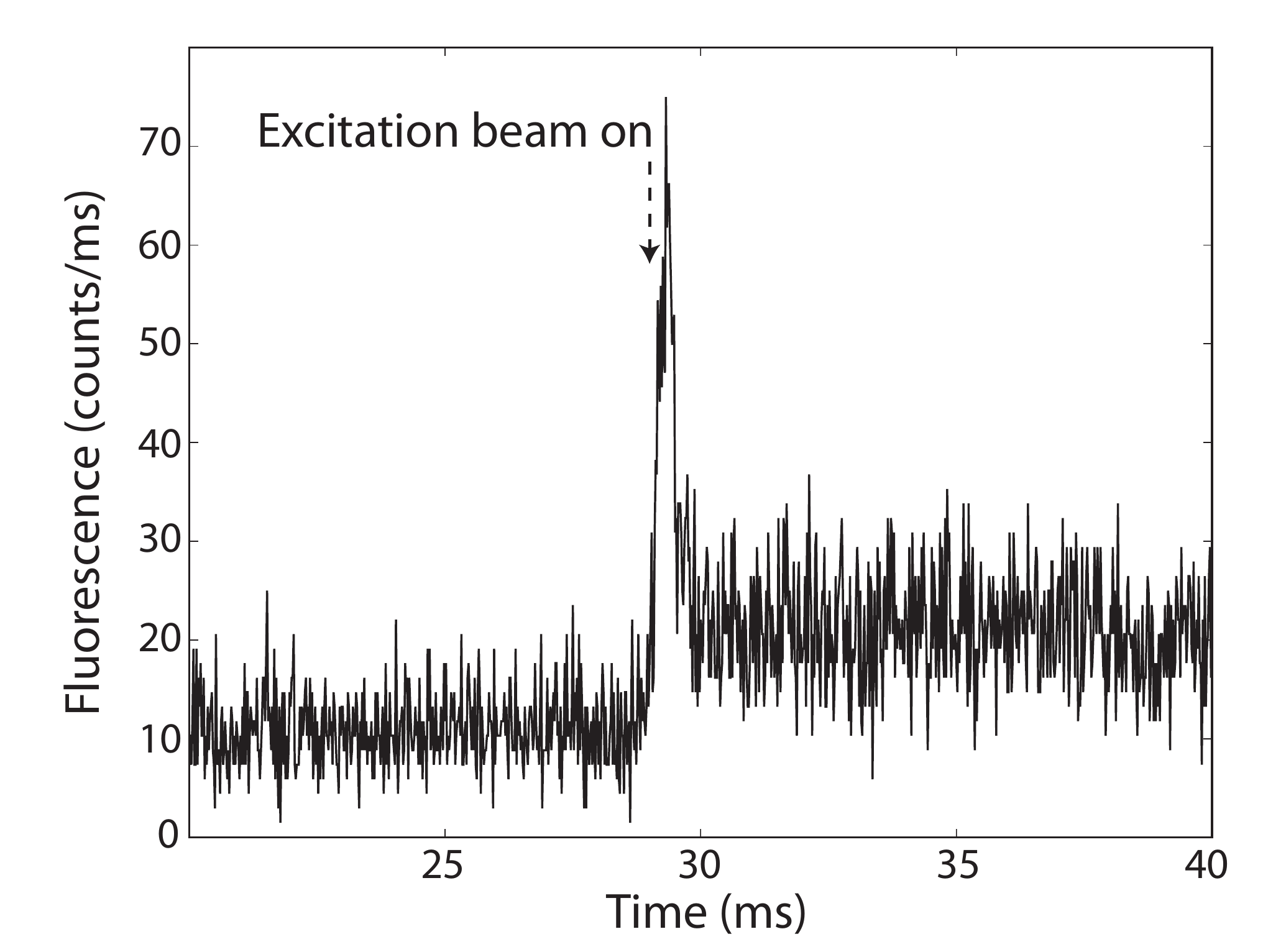}
        \caption{{\bf Atomic fluorescence signal collected by the waveguide.} A laser beam at normal incidence to the chip illuminates atoms in the trench, and generates the fluorescence signal shown in the graph (average of 34 shots). Atoms lying within the $\sim\pi w_0 ^2L$ volume of the waveguide mode, emit a fraction $\sim w_0^2/L^2$ of their fluorescence into the numerical aperture of the waveguide. Roughly one twelfth of these photons produce counts in the avalanche photodiode because of the losses in the fibre connections and quantum efficiency of the APD. The scattering rate per atom is close to the saturated rate of $1.9\times10^7\,$s$^{-1}$ and the atom density is $\rho\sim1\times10^{-2}\,\mu$m$^{-3}$. For these parameters, the estimated fluorescence is 70\,kHz. This estimate is in (unduly) good agreement with our observation, indicating that the fluorescence detection methods works well in our device.}
        \label{Fig: fluorescence}
\end{figure}

It is useful to count the number of atoms in the junction by measuring their fluorescence. We have tested this idea by pushing atoms into the trench, then turning on an intense excitation beam and detecting the fluorescence collected by the waveguide. The expected rate is approximately $\rho R \pi w_0^4 \eta/ L$ where $R=\Gamma/2$ is the scattering rate per atom, $w_0=2.2~\mu\rm{m}$ is the mode size, and $\eta\simeq 1/12$ accounts for transmission loss and detector inefficiency. Fig.~\ref{Fig: fluorescence} shows the fluorescence signal at the APD rising to a peak of $7\times 10^4$ counts per second after the excitation beam is turned on, as expected for number density $\rho=10^{-2}\,\mu$m$^{-3}$. This agreement shows that the waveguide collects more than 1\% of the fluorescence from atoms within the volume of the junction, making this another practical way to count the atoms.

In conclusion, we have reported the fabrication and first demonstration of a 12-channel, micro-fabricated atom-photon junction. We have shown that this chip can detect a low density of atoms, corresponding to less than one atom in the light mode. Conversely, we have shown that atoms in the trench can be used to determine the intensity of the light in the junction. We have also used them to measure the polarisation of the light and to establish that this can be reliably controlled from the input fibre. Our next step will be to bring an ultra-cold cloud of magnetically trapped atoms into the trench. These clouds have densities in the range $1 - 100$~atoms/$\mu$m$^{3}$ and will therefore be opaque at resonance. If off-resonant measurements\cite{Hope05} are used, much less than one photon per atom can be scattered, while retaining sufficient sensitivity to make precise measurements of local density in a single shot as short as 1$\mu$s.

\section*{Methods}
\subsection{Photonic chip fabrication.\label{sec: fabrication}}

The chip was manufactured by the Centre for Integrated Photonics\cite{CIP} on a 1\,mm-thick silicon wafer. A 10\,$\mu$m-thick layer of silica, grown by thermal oxidation, formed the lower cladding for the waveguides. A second layer of silica, doped with Germanium and Boron to achieve a 0.75\% refractive index contrast, was created by flame hydrolysis deposition. This was etched through a UV-lithography mask to create the waveguide cores. A further 10\,$\mu$m silica layer of upper cladding, also deposited by flame hydrolysis, was doped with boron and phosphorus in order to match the refractive index of the lower cladding. A layer of 50\,nm of chrome and then 100\,nm of gold provides the reflecting surface needed by the MOT beams. Finally the central trench was cut by deep reactive ion etching to a depth of 22\,$\mu$m through a 16\,$\mu$m$\times$500\,$\mu$m rectangular mask formed by UV-lithography. The finished chip was polished on the back to reduce its thickness to 500\,$\mu$m. This reduced the distance between the atoms in the trench and the current-carrying wires underneath.

The chip is glued to a sub-chip, containing the current-carrying wires. These provide magnetic fields for manipulating the atoms, for example, the field parallel to the waveguides that is used in the experiments of Fig.~\ref{Fig: zeeman}. Once the chip is attached to this base, a single-mode fibre (non-polarisation-maintaining) is brought to the polished edge of the chip and light is coupled into one of the waveguides, as can be seen by the scatter from defects in the trench. The whole chip is heated to 80~$^\circ$C, and Epotek 353ND index-matching adhesive is placed on the interface between the fibre and the chip. The fibre is realigned repeatedly during the 2-hour curing process, and the joint is reinforced with extra adhesive. We handled each fibre individually by gluing its last 20~mm to a V-groove in a titanium splint. This method allowed us to compensate for variations in concentricity of the fibre core. The procedure was repeated for each waveguide that was connected.

The entire optical path from input fibre to output shows a 5\% transmission. This is made up of 50\% from each of the two vacuum feedthroughs, 60\% from each fibre/waveguide interface, 78\% at each face of the trench and 83\% mode overlap where the expanded beam arrives at the second waveguide. If there is any cross-talk between the waveguides it is less than $10^{-3}$. The various interfaces in the optical train form several etalons that modulate the transmission by a few percent as the frequency varies.

\subsection{Detection system.}
The light is transported by optical fibres to a single-photon counting avalanche photodiode. The counting rates in these experiments are between $7 \times 10^4$ and $1.5\times 10^7$ counts per second, far above the background of $\sim 130\,$s$^{-1}$ from dark counts and stray light. At the high end, almost half the counts are lost to the 32\,ns dead time of the detector, which limits the rate. We extended the dynamic range by using calibrated filters to attenuate the light.  We checked and confirmed the manufacturer's specified dead-time by measuring the noise as a function of intensity and comparing this with the Poisson statistics appropriate for attenuated laser light.

Our measurements of absorption were taken over a number of shots, ranging from 30 to 600. The APD signal was averaged, then smoothed over a 2\,ms window before taking the reading at the point in time where the atom density in the trench was at its maximum. For the polarisation measurements the large magnetic field needed time to settle after switching it on, and therefore we delayed the measurement point. Each of these measurements was accompanied by a background measurement without atoms, which we averaged and used to normalise the absorption. The error bars are the $1\sigma$ statistical uncertainties of a Poissonian photon source with detector dead time \cite{Yu00}.

\subsection{Saturation intensity $I_{sat}$.\label{sec: Isat}}

Light is at the saturation intensity $I_{sat}$ when the photon scattering rate from one atom is $\Gamma/4$. Here, $\Gamma$ is the population decay rate of the upper state and is the same regardless of which magnetic sub-level is excited. Among the  $D_2$ transitions $\left|F=2,m_F\right>\rightarrow\left|F'=3,m_F'\right>$ of $^{87}$Rb, the strongest are $m_F=\pm2 \rightarrow m_F'=\pm3$ driven by $\sigma^\pm$ light. Both of these have $I_{sat}=16.7~\rm{pW}/(\mu\rm{m})^2$. The relative strengths of the other transitions depend only on the squares of the 3-j symbols $\begin{pmatrix}
 3 & 1 & 2 \\ -m_F'  & -m_F'+m_F & m_F
 \end{pmatrix}^2$. A magnetic field is applied along the waveguides, so the light only has $\sigma^\pm$ polarisations and the transitions are restricted to $m_F'=m_F\pm 1$.

For the experiments in Fig.~\ref{Fig: absorption}, the atoms scatter several photons, even in the most weakly excited sub-level, and this ensures that the cloud is optically pumped. To a good approximation, the resulting population distribution of $m_F$ states is the steady-state of the Einstein rate equations. We calculate the mean scattering rate using that distribution. The calculation is repeated to average over all polarisations of the light because our measurements are an average over many shots, each with random polarisation. Thus, we obtain an effective saturation intensity of $I_{sat}=22.4~\rm{pW}/(\mu\rm{m})^2$.
\newline

\paragraph{Acknowledgements:} We acknowledge valuable discussions with Benoit Darqui\'e and the technical expertise of Jon Dyne, Stephen Maine and Valerijus Gerulis, without whom the apparatus could not have been constructed. We acknowledge UK support by EPSRC, QIPIRC and the Royal Society and EU support through SCALA.
\paragraph{Competing Financial Interests:} We the authors declare that there are no competing financial interests.
\paragraph{Author Contributions:} All the authors were involved in building the apparatus and interpreting the data.
\paragraph{Correspondence:} Correspondence should be addressed to Robert Nyman~(email: r.nyman@imperial.ac.uk) or Ed Hinds (ed.hinds@imperial.ac.uk).


\begin{thebibliography}{10}
\bibitem{footnote1}
The fibres were not well fixed for the experiments of Fig.~\ref{Fig: absorption}, where the polarisation was random.
\end{thebibliography}
\end{document}